\documentclass[a4paper,UKenglish]{lipics-v2016}
 
\usepackage{microtype}

\usepackage[sort,nocompress]{cite}

\newcommand{\bigO}{\mathcal{O}}
\newcommand{\DYN}{\texttt{DYNAMIC}}
\newcommand{\runs}{r}

\graphicspath{{./pictures/}}

\title{A Framework of Dynamic Data Structures for String Processing}

\author{Nicola Prezza\thanks{Part of this work was done while the author was a PhD student at the University of Udine, Italy. Work supported by the Danish Research Council (DFF-4005-00267)}}
\affil{Technical University of Denmark, DTU Compute\\ \texttt{npre@dtu.dk}}
\authorrunning{N. Prezza} 

\Copyright{Nicola Prezza}


\EventEditors{John Q. Open and Joan R. Acces}
\EventNoEds{2}
\EventLongTitle{42nd Conference on Very Important Topics (CVIT 2016)}
\EventShortTitle{CVIT 2016}
\EventAcronym{CVIT}
\EventYear{2016}
\EventDate{December 24--27, 2016}
\EventLocation{Little Whinging, United Kingdom}
\EventLogo{}
\SeriesVolume{42}
\ArticleNo{23}

\begin{document}

\maketitle

\begin{abstract}
In this paper we present \DYN, an open-source C++ library implementing dynamic compressed data structures for string manipulation. Our framework includes useful tools such as searchable partial sums, succinct/gap-encoded bitvectors, and entropy/run-length compressed strings and FM-indexes. 
We prove close-to-optimal theoretical bounds for the resources used by our structures, and show that our theoretical predictions are empirically tightly verified in practice. To conclude, we turn our attention to applications. We compare the performance of four recently-published compression algorithms implemented using \DYN~with those of state-of-the-art tools performing the same task. Our experiments show that algorithms making use of dynamic compressed data structures can be up to three orders of magnitude more space-efficient (albeit slower) than classical ones performing the same tasks.
\end{abstract}

\section{Introduction}

Dynamism is an extremely useful feature in the field of data structures for string manipulation, and has been the subject of study in many recent works~\cite{cordova2016practical,klitzke2016general,navarro2014optimal,raman2001succinct,makinen2008dynamic,grossi2013dynamic}. These results showed that---in theory---it is possible to match information-theoretic spatial upper and lower bounds of many problems related to dynamic data structures while still supporting queries in provably optimal time. From the practical point of view however, many of these results are based on too complicated structures which prevent them to be competitive in practice. This is due to several factors that in practice play an important role but in theory are often poorly modeled: cache locality, branch prediction, disk accesses, context switches, memory fragmentation. Good implementations must take into account all these factors in order to be practical. 
Dynamic data structures are based on components that are often cache-inefficient and memory-consuming (e.g. self-balancing trees) and therefore easily run into the above-mentioned problems; 
this is the main reason why little work in this field has been done on the experimental side. An interesting and promising (but still under development) step in this direction is represented by \texttt{Memoria}~\cite{memoria}, a C++14 framework providing general purpose dynamic data structures. Other libraries are also still under development (\texttt{ds-vector}~\cite{ds-vector}) or have been published but the code is not available~\cite{cordova2016practical,klitzke2016general}. To the best of our knowledge, the only working implementation of a dynamic succinct  bitvector is~\cite{gigante-bitvector}.
This situation changes dramatically if the requirement of dynamism is dropped. In recent years, several excellent libraries implementing static data structures have been proposed: \texttt{sdsl}~\cite{gbmp2014sea} (probably the most comprehensive, used, and tested), \texttt{pizza\&chili}~\cite{pizzachili} (compressed indexes), \texttt{sux}~\cite{sux}, \texttt{succinct}~\cite{giuseppe-succinct}, \texttt{libcds}~\cite{libcds}. These libraries proved that \emph{static} succinct data structures can be very practical in addition to being theoretically appealing.

In view of this gap between theoretical and practical advances in the field, in this paper we present \DYN: a C++11 library providing practical implementations of some basic succinct and compressed dynamic data structures for string manipulation: searchable partial sums, succinct/gap-encoded bitvectors, and entropy/run-length compressed strings and FM-indexes. Our library has been extensively profiled and tested, and offers structures whose performance are provably close to the theoretical lower bounds (in particular, they approach succinctness and logarithmic queries). \DYN~is an open-source project and is available at~\cite{prezza2016dynamic}.

We conclude by discussing the performance of four recently-published BWT/LZ77 compression algorithms~\cite{policriti2016computing,policriti2015fast,policriti2015average}  implemented with our library. On highly compressible datasets, our algorithms turn out to be up to three orders of magnitude more space-efficient than classical algorithms performing the same tasks.

\section{The \DYN~library}

The core of our library is a searchable partial sum with inserts data structure (SPSI in what follows). We start by formally defining the SPSI problem and showing how we solve it in \DYN. We then proceed by describing how we use the SPSI structure as building block to obtain the dynamic structures implemented in our library.

\subsection{The Core: Searchable Partial Sums with Inserts}\label{sec: SPSI practice}

The  Searchable Partial Sums With Inserts (SPSI) problem asks for a data structure $ PS $  maintaining a sequence $s_1, \ldots, s_m$ of non-negative $k$-bits integers and supporting the following operations on it:

\begin{itemize}
	\item $\mathtt{PS.sum}(i) = \sum_{j=1}^{i} s_j$;
	\item $\texttt{PS.search}(x)$ is the smallest $i$ such that $\sum_{j=1}^{i}s_j > x$;
	\item $\texttt{PS.update}(i,\delta)$: update $s_i$ to $s_i+\delta$. $\delta$ can be negative as long as $s_i+\delta\geq 0$;
	\item $\texttt{PS.insert}(i)$: insert $0$ between $s_{i-1}$ and $s_i$ (if $i=0$, insert in first position).
\end{itemize}

As discussed later, a consequence of the fact that our SPSI does not support \texttt{delete} operations is that also the structures we derive from it do not support \texttt{delete}; we plan to add this feature in our library in the future.

\DYN's SPSI is a B-tree storing integers  $s_1, \ldots, s_m$ in its leaves and subtree size/partial sum counters in internal nodes. SPSI's operations are implemented by traversing the tree from the root to a target leaf and accessing internal nodes' counters to obtain the information needed for tree traversal. The choice of employing B-trees is motivated by the fact that a big node fanout translates to smaller tree height (w.r.t. a binary tree) and nodes that can fully fit in a cache line (i.e. higher cache efficiency). We use a leaf size $l$ (i.e. number of integers stored in each leaf) always bounded by
$$0.5 \log m \leq l \leq \log m$$
and a node fanout $f\in\bigO(1)$. $f$ has to be chosen accordingly with the cache line size; a bigger value for $f$ reduces cache misses and tree height but increases the asymptotic cost of handling single nodes. See Section \ref{sec:plug n play} for a discussion on the maximum leaf size and $f$ values used in practice in our implementation. 
Letting $l = c\cdot \log m$ being the size of a particular leaf, we call the coefficient $0.5 \leq c \leq 1$ the \emph{leaf load}.

In order to improve space usage even further while still guaranteeing very fast operations, integers in the leaves are packed contiguously in word arrays and, inside each leaf $\mathcal L$, we assign to each integer the bit-size of the largest integer stored in $\mathcal L$. Whenever an integer overflows the maximum size associated to its leaf (after an \texttt{update} operation), we re-allocate space for all integers in the leaf. This operation takes $\bigO(\log m)$ time, so it does not asymptotically increase the cost of \texttt{update} operations. Crucially, in each leaf we allocate space only for the integers actually stored inside it, and re-allocate space for the whole leaf whenever we insert a new integer or we split the leaf. With this strategy, we do not waste space for half-full leaves\footnote{in practice, to speed up operations we allow a small fraction of the leaf to be empty}. Note moreover that, since the size of each leaf is bounded by $\Theta(\log m)$, re-allocating space for the whole leaf at each insertion does not asymptotically slow down \texttt{insert} operations. 

\subsubsection{Theoretical Guarantees}\label{sec:theoretical guarantees}

Let us denote with $m/\log m \leq L \leq 2m/\log m$ the total number of leaves, with $\mathcal L_j$, $0\leq j < L$, the $j$-th leaf of the B-tree (using any leaf order), and with $I\in \mathcal L_j$ an integer belonging to the $j$-th leaf. The total number of bits stored in the leaves of the tree is 
$$
\sum_{0\leq j < L}\ \sum_{I\in \mathcal L_j} max\_bitsize(\mathcal L_j)
$$
where $max\_bitsize(\mathcal L_j) = \max_{I\in \mathcal L_j}\left( bitsize(I) \right)$ is the bit-size of the largest $I\in \mathcal L_j$, and $bitsize(x) = \lfloor\log_2 x\rfloor +1$ is the number of bits required to write number $x$ in binary. The above quantity is equal to 
$$
\sum_{0\leq j < L}\ c_j\cdot\log m\cdot max\_bitsize(\mathcal L_j)
$$
where $0.5 \leq c_j \leq 1$ is the $j$-th leaf load. Since leaves' loads are always upper-bounded by $1$, the above quantity is upper-bounded by
$$
\log m\sum_{0\leq j < L}\ max\_bitsize(\mathcal L_j)
$$
which, in turn, is upper-bounded by 
$$
\log m\sum_{0\leq j < L}\ bitsize\left(\sum_{I\in \mathcal L_j} I\right) \leq \log m \sum_{0\leq j < L} 1+\log_2\left(1+\sum_{I\in \mathcal L_j} I\right)
$$
In the above inequality, we use the upper-bound $bitsize(x) \leq 1 + \log_2(1+x)$ to deal with the case $x=0$. Let $M=m+\sum_{i=1}^{m}s_i = m+\sum_{0\leq j < L} \sum_{I\in \mathcal L_j} I$ be the sum of all integers stored in the structure plus $m$. From the concavity of $\log$ and from $L\leq 2m/\log m$, it can be derived that the above quantity is upper-bounded by

$$
2m\cdot \left(\log(M/m) +\log\log m + 1\right)
$$

To conclude, we store $\bigO(1)$ pointers/counters of $\bigO(\log M)$ bits each per leaf and internal node. We obtain:

\begin{theorem}\label{th: dyn-spsi}
	Let $s_1,\dots,s_m$ be a sequence of $m$ non-negative integers and $M=m+\sum_{i=1}^{m}s_i$. The partial sum data structure implemented in \DYN~takes at most
	$$
	2\cdot m \left(\log(M/m) + \log\log m + \bigO(\log M/\log m)\right)
	$$
	bits of space and supports \texttt{sum}, \texttt{search}, \texttt{update}, and \texttt{insert} operations on the sequence $s_1,\dots,s_m$ in $\bigO(\log m)$ time.
\end{theorem}

In our experiments we observed that---even taking into account memory fragmentation---the bit-size of our dynamic partial sum structure is well approximated by function $1.19\cdot m \left(\log(M/m) + \log\log m + \log M/\log m\right)$. See the experimental section for full details.

\subsection{Plug and Play with Dynamic Structures}\label{sec:plug n play}

The SPSI structure described in the previous section is used as building block to obtain all dynamic structures of our library. In \DYN, the SPSI structure's type name is \texttt{spsi} and is parametrized on 3 template arguments: the leaf type (here, the type \texttt{packed\_vector} is always used\footnote{\texttt{packed\_vector} is simply a packed vector of $k$-bits integers supporting all SPSI operations in linear time}), the leaf size and the node fanout. \DYN~defines two SPSI types with two different combinations of these parameters:

\begin{lstlisting}
typedef spsi<packed_vector,256,16> packed_spsi;
typedef spsi<packed_vector,8192,16> succinct_spsi;
\end{lstlisting}

The reasons for the particular values chosen for the leaf size and node fanout will be discussed later. We use these two types as basic components in the definition our structures. 

\subsubsection{Gap-Encoded Bitvectors}\label{sec:dyn-gap}

\DYN~implements gap-encoded bitvectors using a SPSI to encode gap lengths: bitvector $0^{s_1-1}10^{s_2-1}1 \dots 0^{s_m-1}1$ ($s_i>0$) is encoded with a partial sum on the sequence $s_1, \dots, s_m$. For space reasons, we do not describe how to reduce the gap-encoded bitvector problem to the SPSI problem; the main idea is to reduce bitvector's \texttt{access} and \texttt{rank} to SPSI's \texttt{search}, bitvector's \texttt{select} to SPSI's \texttt{sum}, bitvector's \texttt{insert$_1$} to SPSI's \texttt{insert}, and bitvector's \texttt{insert$_0$}/\texttt{delete$_0$} to SPSI's \texttt{update}.

\DYN's name for the dynamic gap-encoded bitvector class is \texttt{gap\_bitvector}. The class is a template on the SPSI type. We plug \texttt{packed\_spsi} in \texttt{gap\_bitvector} as follows:
\begin{lstlisting}
typedef gap_bitvector<packed_spsi> gap_bv;
\end{lstlisting}
and obtain: 

\begin{theorem}\label{th: dyn-gap-bv}
	Let $B\in\{0,1\}^n$ be a bit-sequence with $b$ bits set. The dynamic gap-encoded bitvector \texttt{gap\_bv} implemented in \DYN~takes at most
	$$
	2\cdot b \left(\log(n/b) + \log\log b + \bigO(\log n/\log b)\right)(1+o(1))
	$$
	bits of space and supports \texttt{rank}, \texttt{select}, \texttt{access}, \texttt{insert}, and \texttt{delete$_0$} operations on $B$ in $\bigO(\log b)$ time.
\end{theorem}

In our experiments, the optimal node fanout for the SPSI stucture employed in this component turned out to be 16, while the optimal leaf size 256 (these values represented a good compromise between query times and space usage).
Our benchmarks show (see the experimental section for full details) that the bit-size of our dynamic gap-encoded bitvector is well approximated by function $1.19\cdot b \left(\log(n/b) + \log\log b + \log n/ \log b\right)$.

\subsubsection{Succinct Bitvectors and Entropy-Compressed Strings}\label{sec:dyn-compressed strings}

Let $n$ be the bitvector length. Dynamic succinct bitvectors can be implemented using a SPSI where all $m=n$ stored integers are either 0 or 1. At this point, \texttt{rank} operations on the bitvector correspond to \texttt{sum} on the partial sum structure, and \texttt{select} operations on the bitvector can be implemented with \texttt{search} on the partial sum structure\footnote{Actually, \texttt{search} permits to implement only \texttt{select$_{1}$}. \texttt{select$_{0}$} can however be easily simulated with the same solution used for \texttt{search} by replacing each integer $x\in\{0,1\}$ with $1-x$ at run time. This solution does not increase space usage.}. \texttt{access} and \texttt{insert} operations on the bitvector correspond to exactly the same operations on the partial sum structure.
Note that in this case we can accelerate operations in the leaves by a factor of $\log n$  by using constant-time built-in bitwise operations such as \texttt{popcount}, masks and shifts. This allows us to use bigger leaves containing $\Theta(\log^2 n)$ bits, which results in a total number of internal nodes bounded by $\bigO(n/\log^2n)$. The overhead for storing internal nodes is therefore of $o(n)$ bits. Moreover, since in the leaves we allocate only the \emph{necessary} space to store the bitvector's content (i.e. we do not allow empty space in the leaves), it easily follows that the dynamic bitvector structure implemented in \DYN~takes $n+o(n)$ bits of space and supports all operations in $\bigO(\log n)$ time. 

In our experiments, the optimal node fanout for the SPSI stucture employed in the succinct bitvector structure turned out to be 16, while the optimal leaf size 8192. \DYN's name for the dynamic succinct bitvector is \texttt{succinct\_bitvector}. The class is a template on the SPSI type. \DYN~defines its dynamic succinct bitvector type as:
\begin{lstlisting}
typedef succinct_bitvector<succinct_spsi> suc_bv;
\end{lstlisting}
We obtain:

\begin{theorem}\label{th: dyn-suc-bitv}
	Let $B\in\{0,1\}^n$ be a bit-sequence. The dynamic succinct bitvector data structure \texttt{suc\_bv} implemented in \DYN~takes  $n+o(n)$ bits of space and supports \texttt{rank}, \texttt{select}, \texttt{access}, and \texttt{insert} operations on $B$ in $\bigO(\log n)$ time. 
\end{theorem}

In our experiments (see the experimental section) the size of our dynamic succinct bitvector was always upper-bounded by $1.23\cdot n$ bits. The $23\%$ overhead on top of the optimal size comes mostly from memory fragmentation ($16\%$). The remaining $7\%$ comes from succinct structures on top of the bit-sequence.

Dynamic compressed strings are implemented with a  wavelet tree built upon dynamic succinct bitvectors. We explicitly store the topology of the tree ($\bigO(|\Sigma|\log n)$ bits) instead of encoding it implicitly in a single bitvector. This choice is space-inefficient for very large alphabets, but reduces the number of \texttt{rank/select} operations on the bitvector(s) with respect of a wavelet tree stored as a single bitvector.
\DYN's compressed strings (wavelet trees) are a template on the bitvector type. \DYN~defines its dynamic string type as:
\begin{lstlisting}
typedef wt_string<suc_bv> wt_str;
\end{lstlisting}

The user can choose at construction time whether to use a Huffman, fixed-size, or Gamma encoding for the alphabet. Gamma encoding is useful when the alphabet size is unknown at construction time. When using Huffman topology, the implementation satisfies:

\begin{theorem}\label{th: dyn-H0-strings}
	Let $S\in\Sigma^n$ be a string with zero-order entropy equal to $H_0$. The Huffman-compressed dynamic string data structure \texttt{wt\_str} implemented in \DYN~takes 
	$$
	n(H_0+1)(1+o(1)) + \bigO(|\Sigma|\log n)
	$$
	bits of space and supports \texttt{rank}, \texttt{select}, \texttt{access}, and \texttt{insert} operations on $S$ in average $\bigO((H_0+1)\log n)$ time. 
\end{theorem}

 In the case a fixed-size encoding is used (i.e. $\lceil\log_2|\Sigma|\rceil$ bits per character), the structure takes $n\log|\Sigma|(1+o(1))+ \bigO(|\Sigma|\log n)$ bits of space and supports all operations in $\bigO(\log|\Sigma|\cdot\log n)$ time.

\subsubsection{Run-Length Encoded Strings}

To run-length encode a string $S\in\Sigma^n$, we adopt the approach described in~\cite{siren2009run}. We store one character per run in a string $H\in \Sigma^r$, we mark the end of the runs with a bit set in a bit-vector $V_{all}[0,\ldots,n-1]$, and for every $ c \in \Sigma $ we store all $c$-runs lengths consecutively in a bit-vector $ V_c $ as follows: every $ m $-length $ c $-run is represented in $ V_c $ as $0^{m-1}1$. 

\begin{example}
	Let $S=bc\#bbbbccccbaaaaaaaaaaa$. We have: $H=bc\#bcba$, $V_{all}=11100010001100000000001$, $V_a=00000000001$, $V_b = 100011$, $V_c = 10001$, and $V_{\#} = 1$
\end{example}

By encoding $H$ with a wavelet tree and gap-compressing all bitvectors, we achieve run-length compression. It can be easily shown that this representation allows supporting \texttt{rank}, \texttt{select}, \texttt{access}, and \texttt{insert} operations on $S$, but for space reasons we do not give these details here. In \DYN, the run-length compressed string type \texttt{rle\_string} is a template on the gap-encoded bitvector type (bitvectors $V_{all}$ and $V_c,\ c\in\Sigma$) and on the dynamic string type (run heads $H$). We plug the structures of the previous sections in the above representation as follows:

\begin{lstlisting}
typedef rle_string<gap_bv, wt_str> rle_str;
\end{lstlisting}
and obtain:

\begin{theorem}\label{th: dyn-rle-string}
	Let $S\in\Sigma^n$ be a string with $\runs_S$ equal-letter runs. The  dynamic run-length encoded string data structure \texttt{rle\_str} implemented in \DYN~takes 
	$$
	\runs_S\cdot \left(4\log(n/\runs_S) + \log|\Sigma| + 4\log\log \runs_S + \bigO(\log n/\log \runs_S)\right)(1+o(1)) + \bigO(|\Sigma|\log n)
	$$
	bits of space and supports \texttt{rank}, \texttt{select}, \texttt{access}, and \texttt{insert} operations on $S$ in $\bigO(\log|\Sigma|\cdot \log \runs_S)$ time. 
\end{theorem}

\subsubsection{Dynamic FM-Indexes}

We obtain dynamic FM-indexes by combining a dynamic Burrows-Wheeler transform with a sparse dynamic vector storing the suffix array sampling. 
In \DYN, the BWT is a template class parametrized on the L-column and F-column types. For the F column, a run-length encoded string is always used. \DYN~defines two types of dynamic Burrows-Wheeler transform structures (wavelet-tree/run-length encoded):

\begin{lstlisting}
typedef bwt<wt_str,rle_str> wt_bwt;
typedef bwt<rle_str,rle_str> rle_bwt;
\end{lstlisting}

Dynamic sparse vectors are implemented inside the FM index class using a dynamic bitvector marking sampled BWT positions and a dynamic sequence of integers (a SPSI) storing non-null values. We combine a Huffman-compressed BWT with a succinct bitvector and a SPSI:
\begin{lstlisting}
typedef fm_index<wt_bwt, suc_bv, packed_spsi> wt_fmi;
\end{lstlisting}
and obtain:

\begin{theorem}\label{dyn: h0-fm-index}
	Let $S\in\Sigma^n$ be a string with zero-order entropy equal to $H_0$, $P\in\Sigma^m$ a pattern occurring $occ$ times in $T$, and $k$ the suffix array sampling rate. The dynamic Huffman-compressed FM-index \texttt{wt\_fmi} implemented in \DYN~takes 
	$$
	n(H_0+2)(1+o(1)) + \bigO(|\Sigma|\log n) + (n/k)\log n
	$$
	bits of space and supports:
	\begin{itemize}
		\item \texttt{access} to BWT characters in average $\bigO((H_0+1)\log n)$ time
		\item \texttt{count} in average $\bigO(m(H_0+1)\log n)$ time
		\item \texttt{locate} in average $\bigO((m+occ\cdot k)(H_0+1)\log n)$ time
		\item text \texttt{left-extension} in average $\bigO((H_0+1)\log n)$ time
	\end{itemize}
	
	If a plain alphabet encoding is used, all $(H_0+1)$ terms are replaced by $\log|\Sigma|$ and times become worst-case.	
\end{theorem}

If, instead, we combine a run-length compressed BWT with a gap-encoded bitvector and a SPSI as follows:

\begin{lstlisting}
typedef fm_index<rle_bwt, gap_bv, packed_spsi> rle_fmi;
\end{lstlisting}
we obtain:

\begin{theorem}
	Let $S\in\Sigma^n$ be a string whose BWT has $\runs$ runs, $P\in\Sigma^m$ a pattern occurring $occ$ times in $T$, and $k$ the suffix array sampling rate. The dynamic run-length compressed FM-index \texttt{rle\_fmi} implemented in \DYN~takes 
	$$
	\runs\cdot \left(4\log(n/\runs) + \log|\Sigma| + 4\log\log \runs + \bigO(\log n/\log \runs)\right)(1+o(1)) + \bigO(|\Sigma|\log n) + (n/k)\log n
	$$
	bits of space and supports:
	\begin{itemize}
		\item \texttt{access} to BWT characters in $\bigO(\log|\Sigma|\cdot\log \runs)$ time
		\item \texttt{count} in  $\bigO(m\cdot\log|\Sigma|\cdot\log \runs)$ time
		\item \texttt{locate} in $\bigO((m +occ\cdot k)(\log|\Sigma|\cdot\log \runs))$ time
		\item text \texttt{left-extension} in $\bigO(\log|\Sigma|\cdot\log \runs)$ time
	\end{itemize}
\end{theorem}

The suffix array sample rate $k$ can be chosen at construction time. 

\section{Experimental Evaluation}

We start by presenting detailed benchmarks of our gap-encoded and succinct bitvectors, standing at the core of all other library's structures. We then turn our attention to applications: we compare the performance of five recently-published compression algorithms implemented with \DYN~against those of state-of-the-art tools performing the same tasks and working in uncompressed space.  All experiments were performed on a \texttt{intel core i7} machine with 12 GB of RAM running Linux Ubuntu 16.04.

\subsection{Benchmarks: Succinct and Gap-Encoded Bitvectors}

\begin{figure}[!tbp]
	\centering
	\begin{minipage}[b]{0.49\textwidth}
		\includegraphics[width=\textwidth]{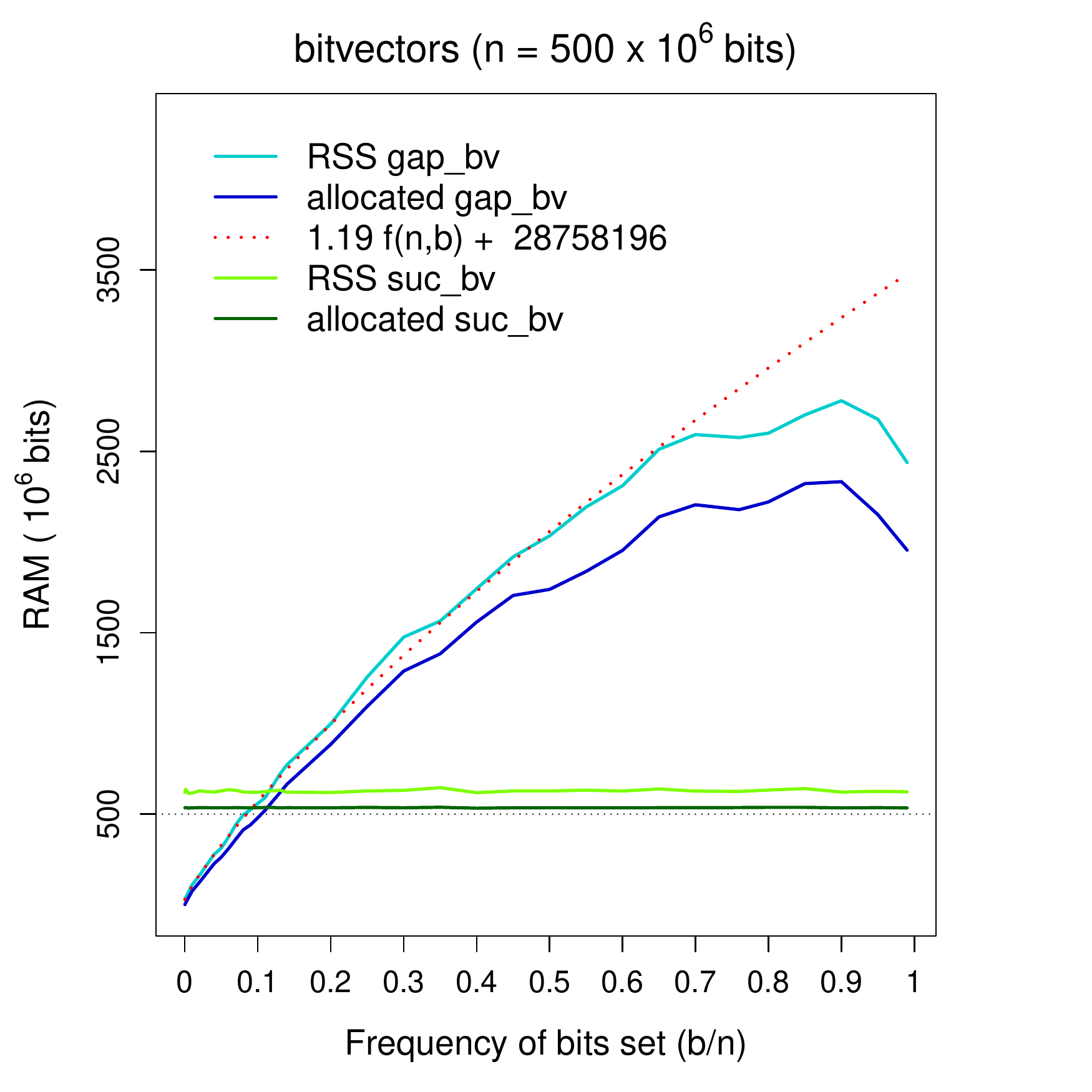}
		\caption{Memory occupancy of \DYN's bitvectors. We show the growth of function $f(n,b) = b(\log (n/b) + \log \log  b+ \log  n/\log  b)$ opportunely scaled to take into account memory fragmentation and the weight of loaded C++ libraries.}\label{fig:bitvectors_space}
	\end{minipage}
	\hfill
	\begin{minipage}[b]{0.49\textwidth}
		\includegraphics[width=\textwidth]{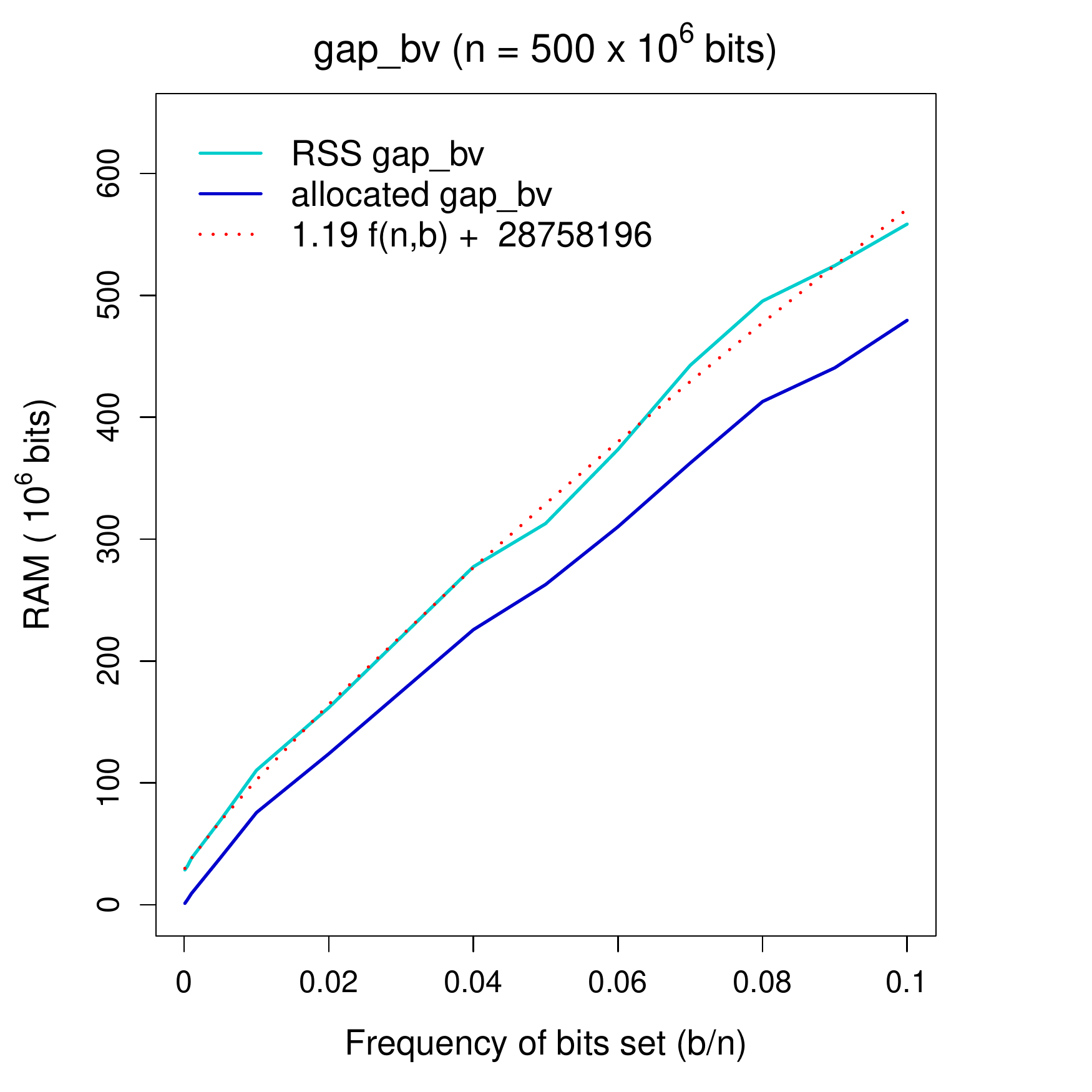}
		\caption{Memory occupancy of \DYN's gap-encoded bitvector in the interval $b/n\in [0,0.1]$ (for $b/n>0.1$ the succinct bitvector is more space-efficient then the gap-encoded one). The picture shows that allocated memory closely follows our theoretical prediction (function $f(n,b)$).}\label{fig:gap_focus_space}
	\end{minipage}
\end{figure}

We built 34 gap-encoded (\texttt{gap\_bv}) and 34 succinct (\texttt{suc\_bv}) bitvectors of length $n=500\cdot 10^6$ bits, varying the frequency $b/n$ of bits set in the interval $[0.0001, 0.99]$. In each experiment, we first built the bitvector by performing $n$ \texttt{insert$_b$} queries, $b$ being equal to 1 with probability $b/n$, at uniform random positions. After building the bitvector, we executed $n$  \texttt{rank$_0$}, $n$  \texttt{rank$_1$}, $n$  \texttt{select$_0$}, $n$  \texttt{select$_1$}, and $n$ \texttt{access} queries at uniform random positions. Running times of each query were averaged over the $n$ repetitions. We measured memory usage in two ways: (i) internally  by counting the total number of bits allocated by our procedures---this value is denoted as \emph{allocated} memory in our plots---, and (ii) externally using the tool \texttt{/usr/bin/time}---this value is denoted as \emph{RSS} in our plots (Resident Set Size).\\\ \\
\textbf{Working space} We fitted measured RSS memory with the theoretical predictions of Section \ref{sec:theoretical guarantees} using a linear regression model. Parameters of the model were inferred using the statistical tool \texttt{R} (function \texttt{lm}). In detail, we fitted RSS memory in the range $b/n\in[0,0.1]$\footnote{For $b/n\geq 0.1$ it becomes more convenient---see below---to use our succinct bitvector, so we considered it more useful to fit memory usage in $b\in[0,0.1]$. In any case---see plot \ref{fig:bitvectors_space}---the inferred model well fits experimental data in the (more wide) interval $b/n\in[0,0.7]$.} with function $k\cdot f(n,b) +c$, where: $f(n,b) = b\cdot (\log (n/b) + \log \log  b+ \log  n/\log  b)$ is our theoretical prediction (recall that memory occupancy of our gap-encoded bitvector should never exceed $2 f(n,b)$), $k$ is a scaling factor accounting for memory fragmentation and  average load distribution in the B-tree, and $c$ is a constant accounting for the weight of loaded C++ libraries (this component cannot be excluded from the measurements of the tool \texttt{/usr/bin/time}). Function \texttt{lm} provided us with parameters $k=1.19$ and $c=28758196$ bits $\approx 3.4 MB$. The value for $c$ was consistent with the space measured with $b/n$ close to 0. 

\begin{figure}[!tbp]
	\centering
	\begin{minipage}[b]{0.49\textwidth}
		\includegraphics[width=\textwidth]{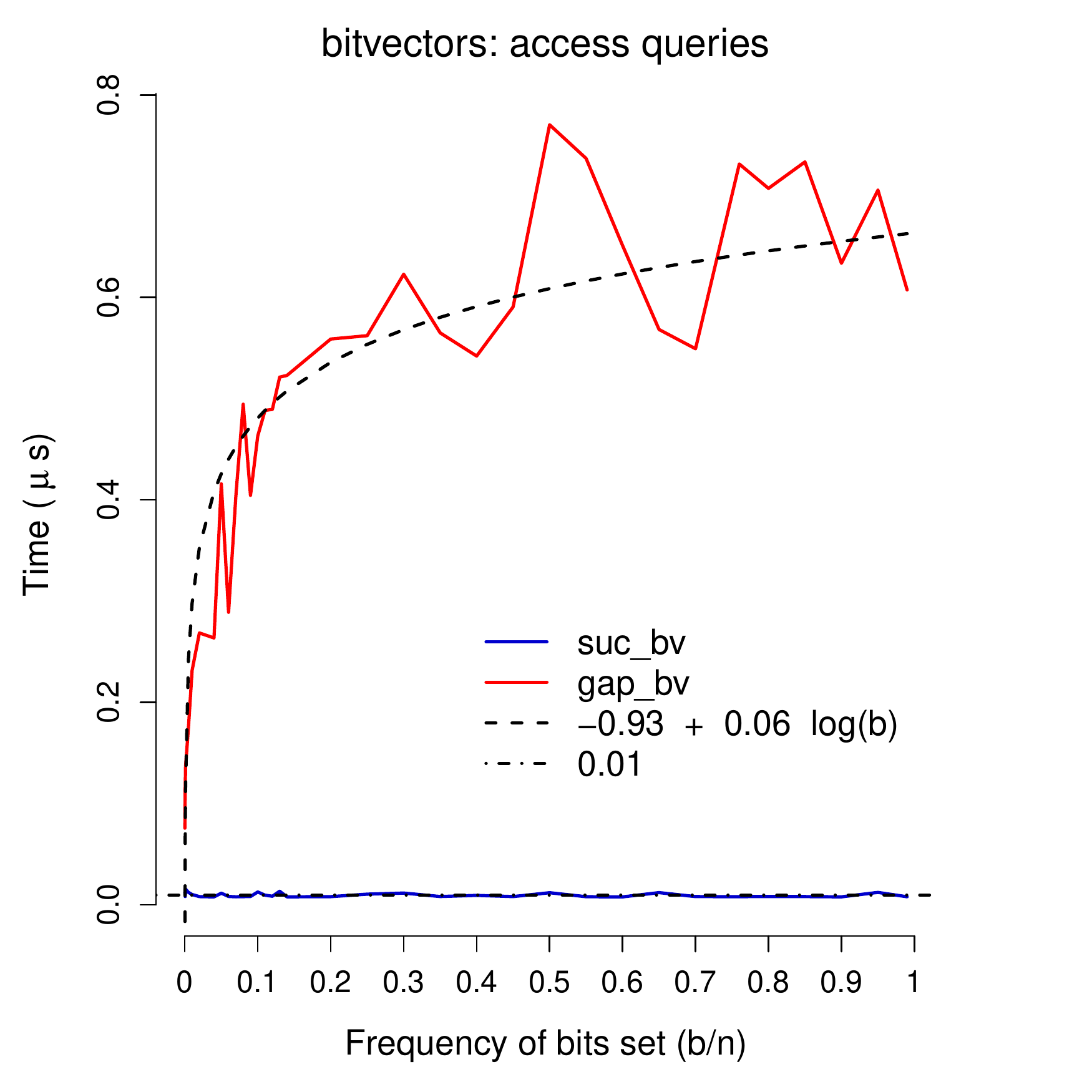}
		\caption{Running times of our bitvectors on \texttt{access} queries. Bitvectors' size is $n=5\times 10^8$ bits.}\label{fig: access}
	\end{minipage}
	\hfill
	\begin{minipage}[b]{0.49\textwidth}
		\includegraphics[width=\textwidth]{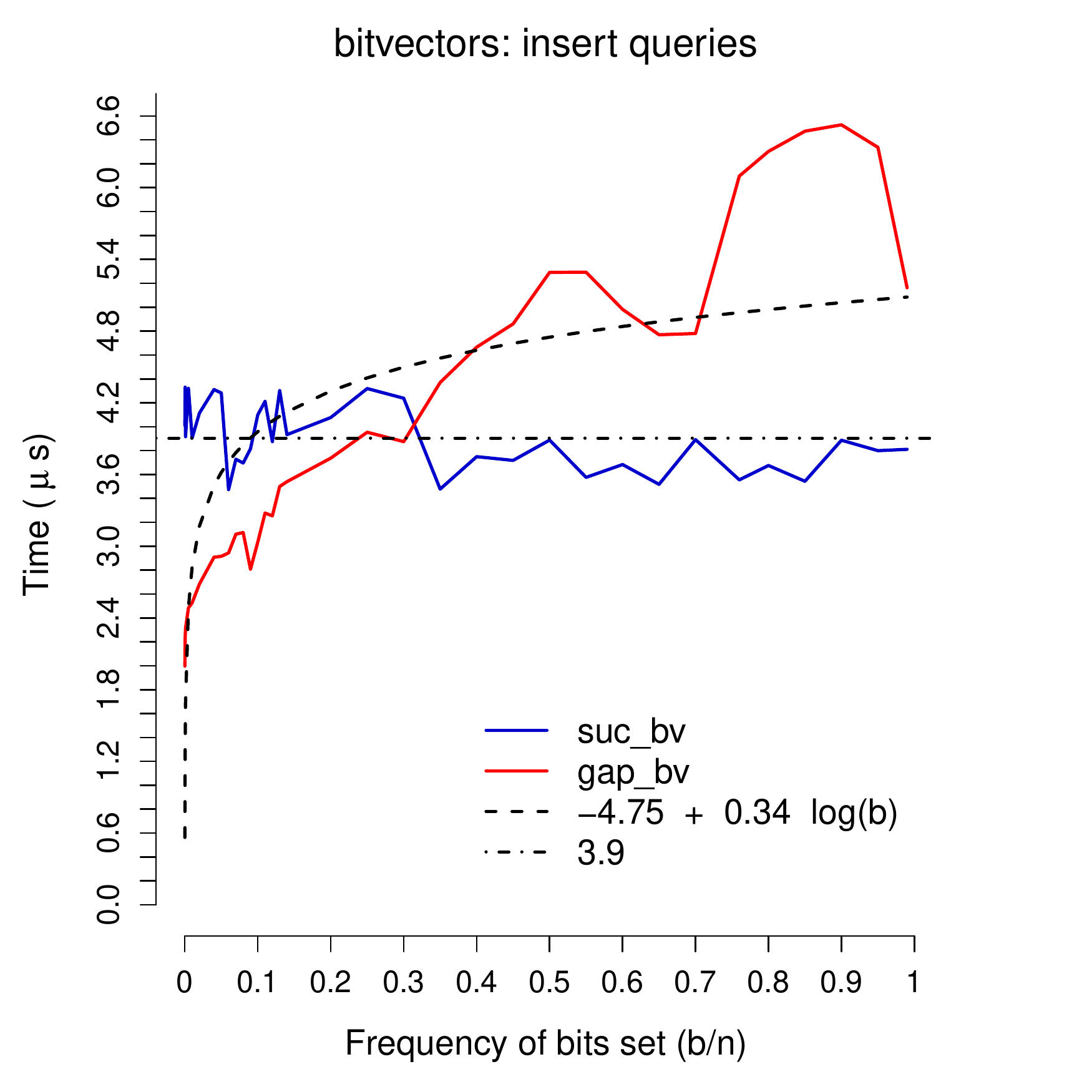}
		\caption{Running times of our bitvectors on \texttt{insert} queries. Bitvectors' size is $n=5\times 10^8$ bits.}\label{fig: insert}
	\end{minipage}
\end{figure}

\begin{figure}[!tbp]
	\centering
	\begin{minipage}[b]{0.49\textwidth}
		\includegraphics[width=\textwidth]{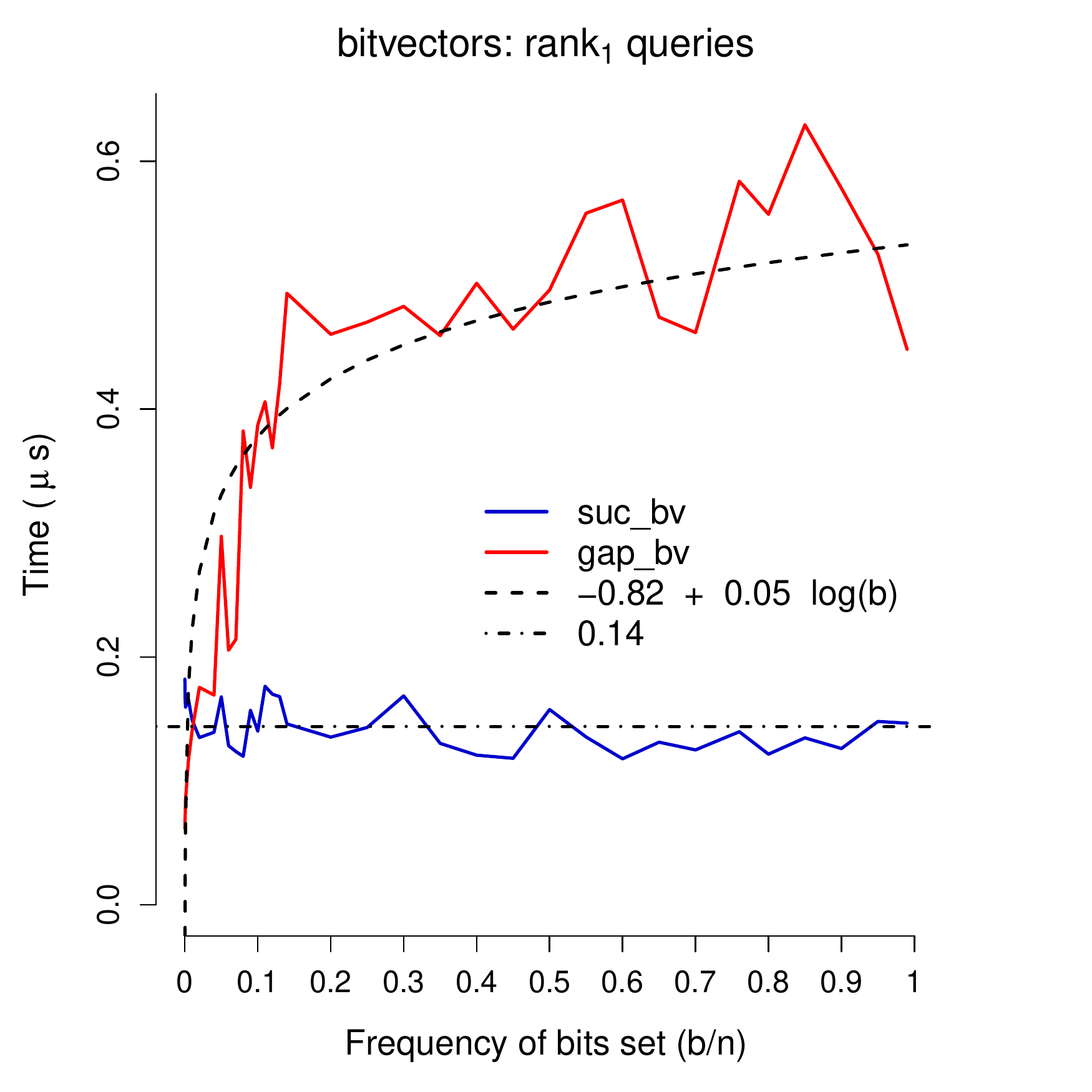}
		\caption{Running times of our bitvectors on \texttt{rank$_1$} queries. Bitvectors' size is $n=5\times 10^8$ bits.}\label{fig: rank1}
	\end{minipage}
	\hfill
	\begin{minipage}[b]{0.49\textwidth}
		\includegraphics[width=\textwidth]{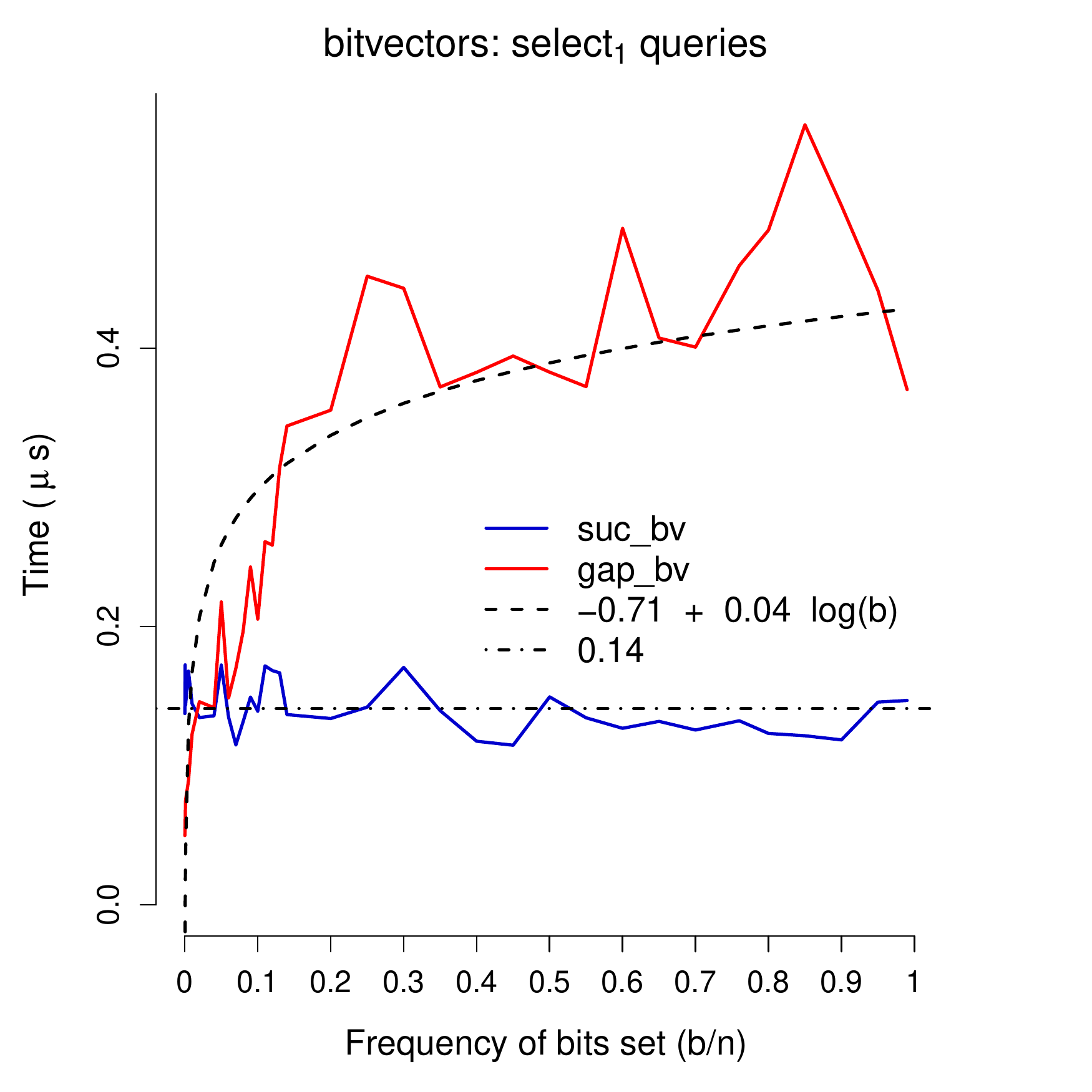}
		\caption{Running times of our bitvectors on \texttt{select$_1$} queries. }\label{fig: select1}
	\end{minipage}
\end{figure}

Figures \ref{fig:bitvectors_space} and \ref{fig:gap_focus_space} show memory occupancy of \DYN's bitvectors as a function of the frequency $b/n$ of bits set. In Figure \ref{fig:bitvectors_space} we compare both bitvectors. In Figure  \ref{fig:gap_focus_space} we focus on the behavior of our gap-encoded bitvector in the interval $b/n\in [0,0.1]$.  In these plots we moreover show the growth of function $1.19\cdot f(n,b) +28758196$. Plot in Figure \ref{fig:bitvectors_space} shows that our theoretical prediction fits almost perfectly the memory usage of our gap-encoded bitvector for $b/n \leq 0.7$. The plot suggests moreover that for $b/n \geq 0.1$ it is preferable to use our succinct bitvector rather than the gap-encoded one.
As far as the gap-encoded bitvector is concerned, memory fragmentation\footnote{we estimated the impact of memory fragmentation by comparing RSS and allocated memory, after subtracting from RSS the estimated weight---approximately $3.4$ MB---of loaded C++ libraries} amounts to approximately $15\%$ of the allocated memory for $b/n\leq 0.5$. This fraction increases to $24\%$ for $b/n$ close to 1. 
We note that RSS memory usage of our succinct bitvector never exceeds $1.29n$ bits: the overhead of $0.29n$ bits is distributed among (1) \texttt{rank/select} succinct structures ($\approx 0.07n$ bits)  (2) loaded C++ libraries (a constant amounting to approximately $3.4$ MB, i.e. $\approx 0.06n$ bits in this case), and memory fragmentation ($\approx 0.16n$ bits). Excluding the size of C++ libraries (which is constant), our bitvector's size never exceeds $1.23n$ bits (being $1.20n$ bits on average).\\\ \\
\textbf{Query times} Plots in Figures \ref{fig: access}-\ref{fig: select1} show running times of our bitvectors on all except \texttt{rank$_0$} and \texttt{select$_0$} queries (which were very close to those of \texttt{rank$_1$} and \texttt{select$_1$} queries, respectively).
We used a linear regression model (inferred using \texttt{R}'s function \texttt{lm}) to fit query times of our gap-encoded bitvector with function $c+ k\cdot \log b$. Query times of our succinct bitvector were interpolated with a constant (being $n$ fixed). These plots show interesting results. First of all, our succinct bitvector supports extremely fast ($0.01 \mu s$ on average) \texttt{access} queries. \texttt{rank} and \texttt{select} queries are, on average, 15 times slower than \texttt{access} queries. As expected, \texttt{insert} queries are very slow, requiring---on average---390 times the time of \texttt{access} queries and 26 times that of \texttt{rank/select} queries. On all except \texttt{access} queries, running times of our gap-encoded bitvector are faster than (or comparable to) those of our succinct bitvector for $b/n\leq 0.1$. Combined with the results depicted in Plot \ref{fig:bitvectors_space}, these considerations confirm that for $b/n \leq 0.1$ our gap-encoded bitvector should be preferred to the succinct one. \texttt{access}, \texttt{rank}, and \texttt{select} queries are all supported in comparable times on our gap-encoded bitvector ($\approx 0.05\cdot \log b\ \mu s$), and are one order of magnitude faster than \texttt{insert} queries.

\subsection{An Application: Space-Efficient Compression Algorithms}

We used \DYN~to implement five recently-published algorithms~\cite{policriti2016computing,policriti2015fast,policriti2015average} computing the Burrows-Wheeler transform~\cite{burrows1994block} (BWT) and the Lempel-Ziv 77 factorization~\cite{ziv1977universal} (LZ77) within compressed working space: \texttt{cw-bwt}~\cite{policriti2015average}  builds a BWT within $n(H_k+1) + o(n\log\sigma)$ bits of working space by breaking it in contexts and encoding each context with a zero-order compressed string; \texttt{rle-bwt} builds the BWT within $\Theta(r)$ words of working space using the structure of Theorem \ref{th: dyn-rle-string}; \texttt{h0-lz77}~\cite{policriti2015fast} computes LZ77 online within $n(H_0+2) + o(n\log\sigma)$ bits using a dynamic zero-order compressed FM-index; \texttt{rle-lz77-1} and \texttt{rle-lz77-2}~\cite{policriti2016computing} build LZ77 within $\Theta(r)$ words of space by employing a run-length encoded BWT augmented with a suffix array sampling based on BWT equal-letter runs and LZ77 factors, respectively. Implementations of these algorithms can be found within the \DYN~library~\cite{prezza2016dynamic}.
We compared running times and working space of our algorithms against those of less space-efficient (but faster) state-of-the-art tools solving the same problems. BWT construction tools: \texttt{se-sais}~\cite{beller2013space,gbmp2014sea} ($\Theta(n)$ Bytes of working space), \texttt{divsufsort}~\cite{mori2005short,gbmp2014sea} ($\Theta(n)$ words), 
\texttt{bwte}~\cite{ferragina2012lightweight} (constant user-defined working space; we always used 256 MB),
\texttt{dbwt}~\cite{dbwt} ($\Theta(n)$ Bytes). LZ77 factorization tools:   \texttt{isa6r}~\cite{kempa2013lempel,helsinkiFactAlgo} ($\Theta(n)$ words), 	\texttt{kkp1s}~\cite{karkkainen2013linear,helsinkiFactAlgo}	($\Theta(n)$ words), \texttt{lzscan}~\cite{karkkainen2013lightweight,helsinkiFactAlgo} ($\Theta(n)$ Bytes). 
We generated two highly repetitive text collections by downloading all versions of the \emph{Boost} library (\url{github.com/boostorg/boost}) and all versions of the English \emph{Einstein}'s Wikipedia page (\url{en.wikipedia.org/wiki/Albert_Einstein}). Both datasets were truncated to $5\cdot10^8$ Bytes to limit RAM usage of the and computation times of the tested tools. The sizes of the 7-Zip-compressed datasets (\url{www.7-zip.org}) were 120 KB (Boost) and 810 KB (Einstein). The datasets can be found within the \DYN~library~\cite{prezza2016dynamic}  (folder \texttt{/datasets/}). RAM usage and running times of the tools were measured using the executable \texttt{/usr/bin/time}.

\begin{figure}[h!]
	\centering
	\includegraphics[width=0.7\textwidth]{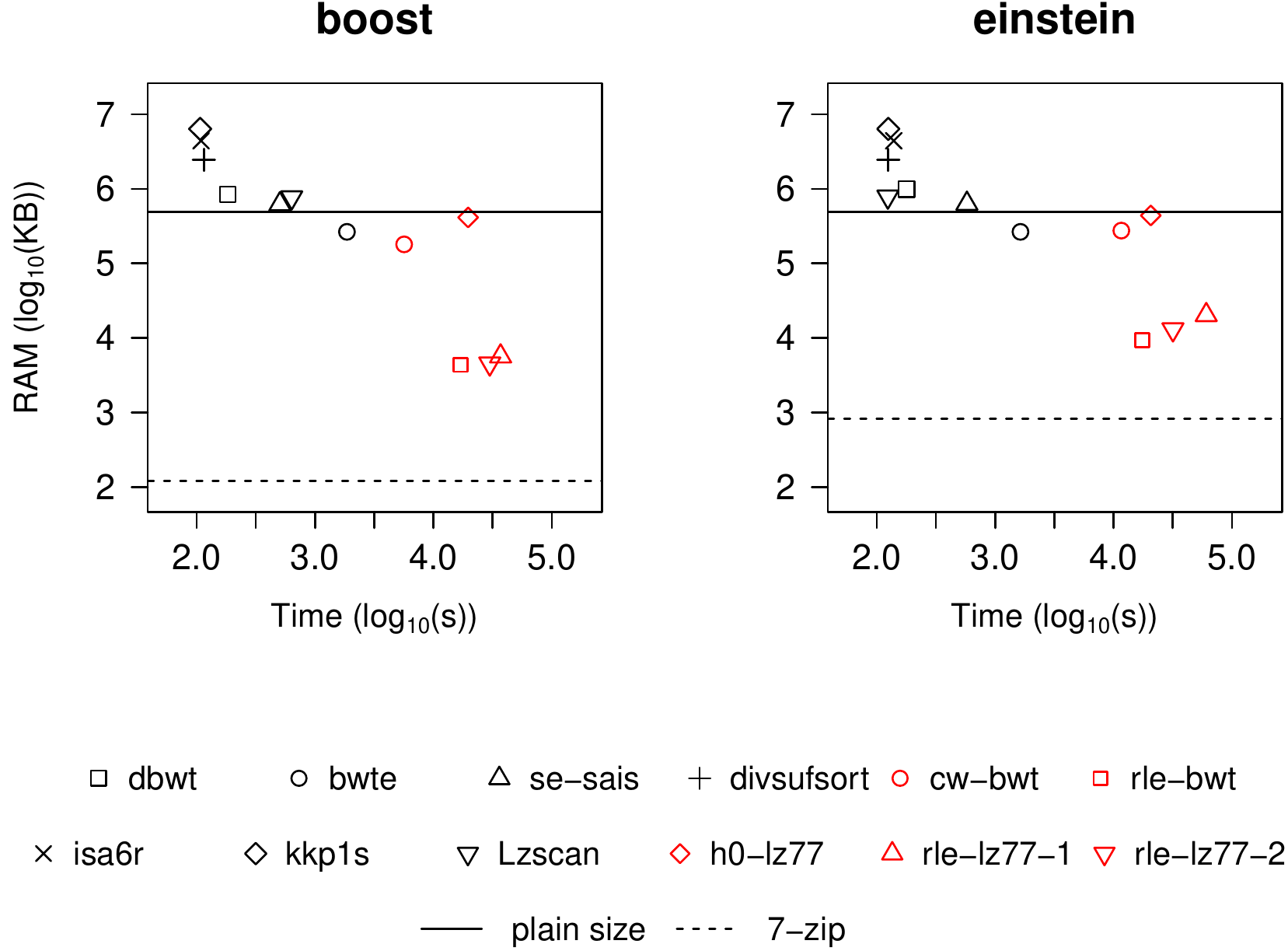}
	\caption{BWT and LZ77 compression algorithms. In red: tools implemented using \DYN. Solid/dashed lines: space of the input files before and after 7-Zip compression, respectively. }\label{fig:compress}
\end{figure}

In Figure \ref{fig:compress} we report our results. Solid and a dashed horizontal lines show the datasets' sizes before and after compression with 7-Zip, respectively. Our tools are highlighted in red. 
We can infer some general trends from the plots. Our tools use always less space than the plain text, and from one to three orders of magnitude more space than the 7-Zip-compressed text. \texttt{h0-lz77} and \texttt{cw-bwt} (entropy compression) use always a working space very close to (and always smaller than) the plain text, with \texttt{cw-bwt} ($k$-th order compression) being more space-efficient than \texttt{h0-lz77} ($0$-order compression). On the other hand, tools using a run-length compressed BWT---\texttt{rle-bwt},  \texttt{rle-lz77-1}, and \texttt{rle-lz77-2}---are up to two orders of magnitude more space-efficient than \texttt{h0-lz77} and \texttt{cw-bwt} in most of the cases. This is a consequence of the fact that run-length encoding of the BWT is particularly effective in compressing repetitive datasets. 
\texttt{bwte} represents a good trade-off in both running times and working space between tools working in compressed and uncompressed working space. \texttt{kkp1s} is the fastest tool, but uses a working space that is one order of magnitude larger than the uncompressed text and three orders of magnitude  larger than that of \texttt{rle-bwt}, \texttt{rle-lz77-1}, and \texttt{rle-lz77-2}. As predicted by theory, tools working in compact working space (\texttt{lzscan}, \texttt{se-sais}, \texttt{dbwt}) use always slightly more space than the uncompressed text, and one order of magnitude less space than tools working in $\bigO(n)$ words. 
To conclude, the plots show that the price to pay for using complex dynamic data structures is high running times: our tools are up to three orders of magnitude slower than tools working in  $\Theta(n)$ words of space. This is mainly due to the large number of \texttt{insert} operations---one per text character---performed by our algorithms to build the dynamic FM indexes.

\vspace{-15pt}



\bibliography{dynamic}

\begin{thebibliography}{10}

\bibitem{beller2013space}
Timo Beller, Maike Zwerger, Simon Gog, and Enno Ohlebusch.
\newblock Space-efficient construction of the burrows-wheeler transform.
\newblock In {\em String Processing and Information Retrieval}, pages 5--16.
  Springer, 2013.

\bibitem{burrows1994block}
Michael Burrows and David~J Wheeler.
\newblock A block-sorting lossless data compression algorithm, 1994.

\bibitem{cordova2016practical}
Joshimar Cordova and Gonzalo Navarro.
\newblock Practical dynamic entropy-compressed bitvectors with applications.
\newblock In {\em International Symposium on Experimental Algorithms}, pages
  105--117. Springer, 2016.

\bibitem{dbwt}
{dbwt}: direct construction of the bwt.
\newblock \url{http://researchmap.jp/muuw41s7s-1587/#_1587}.
\newblock Accessed: 2016-11-17.

\bibitem{ds-vector}
{ds-vector: C++ library for dynamic succinct vector}.
\newblock \url{https://code.google.com/archive/p/ds-vector/}.
\newblock Accessed: 2016-11-17.

\bibitem{prezza2016dynamic}
{DYNAMIC}: dynamic succinct/compressed data structures library.
\newblock \url{https://github.com/xxsds/DYNAMIC}.
\newblock Accessed: 2017-01-22.

\bibitem{ferragina2012lightweight}
Paolo Ferragina, Travis Gagie, and Giovanni Manzini.
\newblock Lightweight data indexing and compression in external memory.
\newblock {\em Algorithmica}, 63(3):707--730, 2012.

\bibitem{gigante-bitvector}
{bitvector: succinct dynamic bitvector implementation}.
\newblock \url{https://github.com/nicola-gigante/bitvector}.
\newblock Accessed: 2016-11-17.

\bibitem{gbmp2014sea}
Simon Gog, Timo Beller, Alistair Moffat, and Matthias Petri.
\newblock From theory to practice: Plug and play with succinct data structures.
\newblock In {\em 13th International Symposium on Experimental Algorithms, (SEA
  2014)}, pages 326--337, 2014.

\bibitem{grossi2013dynamic}
Roberto Grossi, Rajeev Raman, Satti~Srinivasa Rao, and Rossano Venturini.
\newblock Dynamic compressed strings with random access.
\newblock In {\em International Colloquium on Automata, Languages, and
  Programming}, pages 504--515. Springer, 2013.

\bibitem{karkkainen2013lightweight}
Juha K{\"a}rkk{\"a}inen, Dominik Kempa, and Simon~J Puglisi.
\newblock Lightweight lempel-ziv parsing.
\newblock In {\em Experimental Algorithms}, pages 139--150. Springer, 2013.

\bibitem{karkkainen2013linear}
Juha K{\"a}rkk{\"a}inen, Dominik Kempa, and Simon~J Puglisi.
\newblock Linear time {L}empel-{Z}iv factorization: Simple, fast, small.
\newblock In {\em Combinatorial Pattern Matching}. Springer, 2013.

\bibitem{kempa2013lempel}
Dominik Kempa and Simon~J Puglisi.
\newblock Lempel-ziv factorization: Simple, fast, practical.
\newblock In {\em Proceedings of the Meeting on Algorithm Engineering \&
  Expermiments}, pages 103--112. Society for Industrial and Applied
  Mathematics, 2013.

\bibitem{klitzke2016general}
Patrick Klitzke and Patrick~K Nicholson.
\newblock A general framework for dynamic succinct and compressed data
  structures.
\newblock {\em Proceedings of the 18th ALENEX}, pages 160--173, 2016.

\bibitem{libcds}
{libcds: compact data structures library}.
\newblock \url{https://github.com/fclaude/libcds}.
\newblock Accessed: 2016-11-17.

\bibitem{helsinkiFactAlgo}
Lz77 factorization algorithms.
\newblock \url{https://www.cs.helsinki.fi/group/pads/lz77.html}.
\newblock Accessed: 2016-05-20.

\bibitem{makinen2008dynamic}
Veli M{\"a}kinen and Gonzalo Navarro.
\newblock Dynamic entropy-compressed sequences and full-text indexes.
\newblock {\em ACM Transactions on Algorithms (TALG)}, 4(3):32, 2008.

\bibitem{memoria}
{Memoria: C++14 framework providing general purpose dynamic data structures}.
\newblock \url{https://bitbucket.org/vsmirnov/memoria/wiki/Home}.
\newblock Accessed: 2016-11-17.

\bibitem{mori2005short}
Y~Mori.
\newblock Short description of improved two-stage suffix sorting algorithm,
  2005.

\bibitem{navarro2014optimal}
Gonzalo Navarro and Yakov Nekrich.
\newblock Optimal dynamic sequence representations.
\newblock {\em SIAM Journal on Computing}, 43(5):1781--1806, 2014.

\bibitem{pizzachili}
{Pizza\&Chili corpus}.
\newblock \url{http://pizzachili.dcc.uchile.cl}.
\newblock Accessed: 2016-07-25.

\bibitem{policriti2015average}
Alberto Policriti, Nicola Gigante, and Nicola Prezza.
\newblock {Average linear time and compressed space construction of the
  Burrows-Wheeler transform}.
\newblock In {\em International Conference on Language and Automata Theory and
  Applications}, pages 587--598. Springer, 2015.

\bibitem{policriti2015fast}
Alberto Policriti and Nicola Prezza.
\newblock {Fast online Lempel-Ziv factorization in compressed space}.
\newblock In {\em International Symposium on String Processing and Information
  Retrieval}, pages 13--20. Springer, 2015.

\bibitem{policriti2016computing}
Alberto Policriti and Nicola Prezza.
\newblock {Computing LZ77 in run-compressed space}.
\newblock In {\em Data Compression Conference (DCC), 2016}, pages 23--32. IEEE,
  2016.

\bibitem{raman2001succinct}
Rajeev Raman, Venkatesh Raman, and S~Srinivasa Rao.
\newblock Succinct dynamic data structures.
\newblock In {\em Workshop on Algorithms and Data Structures}, pages 426--437.
  Springer, 2001.

\bibitem{siren2009run}
Jouni Sir{\'e}n, Niko V{\"a}lim{\"a}ki, Veli M{\"a}kinen, and Gonzalo Navarro.
\newblock {Run-length compressed indexes are superior for highly repetitive
  sequence collections}.
\newblock In {\em String Processing and Information Retrieval}, pages 164--175.
  Springer, 2009.

\bibitem{giuseppe-succinct}
{succinct library}.
\newblock \url{https://github.com/ot/succinct}.
\newblock Accessed: 2016-11-17.

\bibitem{sux}
{sux library}.
\newblock \url{http://sux.di.unimi.it/}.
\newblock Accessed: 2016-11-17.

\bibitem{ziv1977universal}
Jacob Ziv and Abraham Lempel.
\newblock A universal algorithm for sequential data compression.
\newblock {\em IEEE Transactions on information theory}, 23(3):337--343, 1977.

\end{thebibliography}

\end{document}